# Reproducible Ultrahigh Electromagnetic SERS Enhancement in Nanosphere-Plane Junctions

**Jing Long[1], Hui Yi[1], Hongquan Li[1,2] and Tian Yang[1,★]**

[1]University of Michigan - Shanghai Jiao Tong University Joint Institute, State Key Laboratory of Advanced Optical Communication Systems and Networks, Key Laboratory for Thin Film and Microfabrication of the Ministry of Education, Shanghai Jiao Tong University, Shanghai 200240, China, [2]Currently with Department of Electrical Engineering, Stanford University, Stanford, California 94305, USA. [★]Email: tianyang@sjtu.edu.cn

Surface enhanced Raman scattering (SERS) in nanoscale hotspots has been placed great hopes upon for identification of minimum chemical traces and in-situ investigation of single molecule structures and dynamics[1-7]. However, previous work consists of either irreproducible enhancement factors (EF) from random aggregates, or moderate EFs despite better reproducibility. Consequently, systematic study of SERS at the single and few molecules level is still very limited, and the promised applications are far from being realized. Here we report EFs as high as the most intense hotspots in previous work yet achieved in a reproducible and well controlled manner, that is, electromagnetic EFs (EMEF) of $10^{9\sim10}$ with an error down to $10^{\pm0.08}$ from gold nanospheres on atomically flat gold planes under radially polarized (RP) laser excitation. In addition, our experiment reveals the EF's unexpected nonlinearity under as low as hundreds of nanowatts of laser power.

In the last decade, SERS detections of single small molecules in aggregates of metallic nanoparticles have been confirmed by the bi-analyte method, despite the extreme randomness of hotspot intensities and EFs[8,9]. It has been shown that the EFs vary from around $10^4$ to over $10^{10}$, the 0.0003% most intense hotspots contributing 7% of the overall SERS signal[10]. It has also been pointed out that the most intense hotspots are required for detection of single small molecules with non-resonant Raman scattering cross sections as small as $10^{-29}$ to $10^{-30}$ cm$^2$ sr$^{-1}$ [11]. To resolve the extreme randomness of EFs so as to achieve efficient and systematic study of molecular dynamics, well-controlled fabrication of SERS substrates has been studied extensively[12-16]. For example, electron beam writing of sub-5 nm gap optical antennas has been demonstrated recently, which nevertheless is no longer reproducible at such a small length scale[12,15,16]. An alternative approach that is directly related to our work in this paper is the nanoparticle-plane junction[2,5,17,18], with reported experimental SERS EFs limited to about $10^{6-8}$, and with its reproducibility shown only after averaging tens of hotspots. Meanwhile, there has been great progress in tip-enhanced Raman scattering (TERS) in recent years[6,19-22]. But due to its inherent limited EFs, TERS has only been used to detect single molecules with large Raman scattering cross sections.

In this paper, we report another kind of SERS experiment to achieve both reproducible and ultrahigh SERS EFs for the first time, as shown in Fig. 1a. A chemically synthesized 60 nm gold nanosphere is on top of a 200 nm thick atomically flat gold plane, and a RP He-Ne laser beam at 633 nm is focused by an objective with a numerical aperture (NA) of 0.9 to excite the nanosphere[23]. A monolayer of malachite green isothiocyanate (MGITC) molecules is coated on the surface of the nanosphere whose Raman scattering is collected by the same focusing objective. The nanosphere pairs with its mirror image to form a vertically oriented and vertically polarized optical antenna. As shown in Fig. 1b, the localized surface plasmon resonance (LSPR) spectrum of one of the antennas is measured by collecting its side scattering of a supercontinuum source focused through the objective. The laser wavelength and three of the strongest Raman peaks of MGITC at 1180, 1370 and 1618 cm$^{-1}$ are labeled to show that they all fall within the LSPR resonance. In Fig. 1c, finite difference time domain (FDTD) simulation shows a $7 \times 10^4$ fold increase of the vertical electric field intensity, $|E_z^2|$, in the junction gap hotspot on LSPR resonance, according to classical electromagnetics. A gap height of 1.2 nm and a nominal refractive index of 1.5 have been used in the simulation to represent a monolayer of MGITC on top of the gold plane[24]. The simulated hotspot intensity spectrum is close to the experimental scattering spectrum and shown in Supplementary Fig. S1.

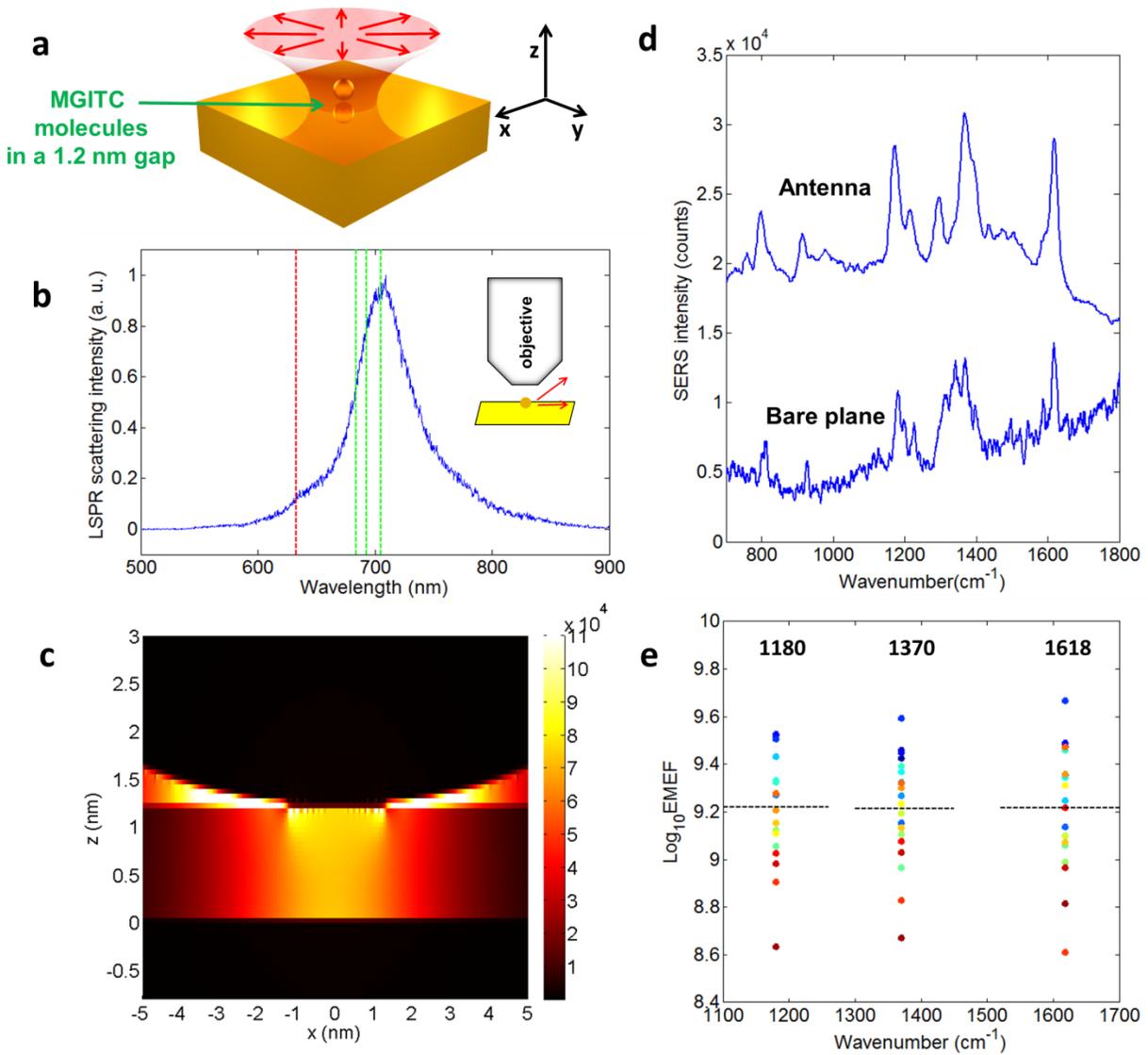

**Figure 1 | LSPR and SERS of antennas coated with a monolayer of MGITC. a,** Illustration of the gold nanosphere-plane antenna under RP excitation. The mirror image of the nanosphere is also plotted. **b,** The LSPR spectrum of an antenna, by measuring its scattering outside of the NA of the illuminating objective. The nanosphere's diameter is 60 nm. The laser wavelength and the three strongest Raman bands in the SERS experiment are labeled as red and green lines, respectively. **c,** FDTD simulation of $|E_z^2|$ in the antenna's junction gap, normalized by the $|E_z^2|$ of an incident p-wave at its resonance wavelength 691 nm. **d,** The SERS spectrum of an antenna, and that of a monolayer of MGITC on a bare gold plane. The laser power at sample is 300 nW for the antenna, and 1.5 mW for the bare plane. The integration time is 4 s for the antenna, and 10 s for the bare plane. **e,**

SERS EMEFs of twenty antennas for three Raman bands at 1180 cm⁻¹, 1370 cm⁻¹ and 1618 cm⁻¹. Each three dots with the same color come from one same antenna. The dashed lines are the average EMEFs for each band.

Figure 1d shows the SERS spectrum of an antenna under a laser power of 300 nW at sample and an integration time of 4 s. To calculate the EF, the Raman spectrum from a monolayer of MGITC coated on a bare atomically flat gold plane under the same RP laser focal spot is also shown. The EMEFs of twenty different antennas for three of the strongest Raman peaks are plotted in Fig. 1e. The EF is defined by comparing with an imaginary experiment in which the molecule is measured in air using a linearly polarized (LP) laser beam and the same focusing objective. Details of EF calculation are described in Supplementary Methods. In the calculation, the hotspot area, $A_{hotspot}$, is taken to be 9.3 nm² according to the FDTD result, which will be discussed later. We attribute the calculated EF to electromagnetic effects, since in both the antenna experiment and the bare gold plane experiment, the thiol group (-SH) of MGITC forms a covalent bond with the gold surface so that they are expected to have chemical EFs close to each other.

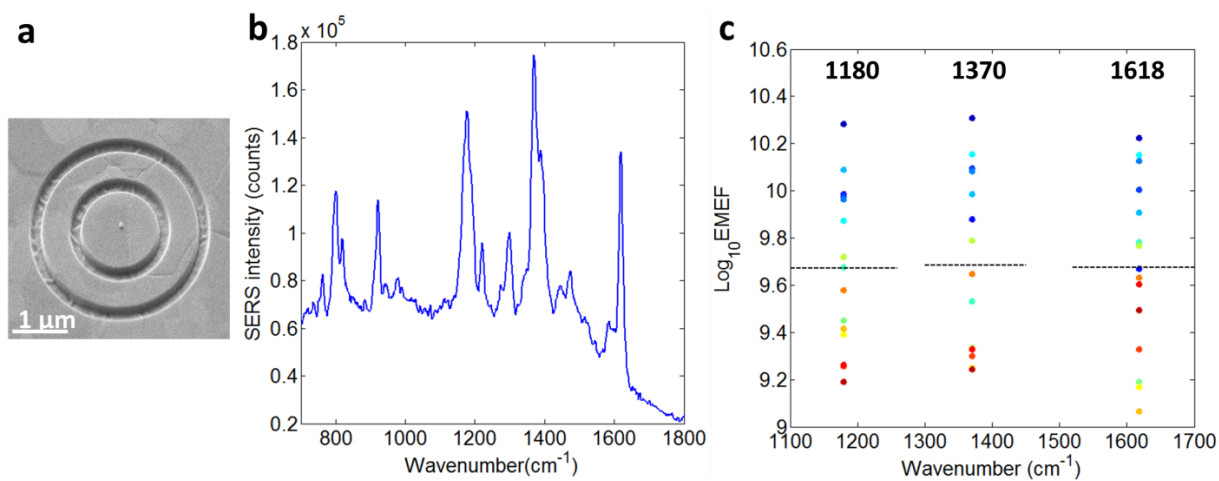

**Figure 2 | SERS of antennas coated with a monolayer of MGITC and enclosed by double rings**. **a**, A scanning electron micrograph of the device. **b**, A typical SERS spectrum. The laser power at sample is 300 nW. The

integration time is 4 s. **c,** SERS EMEFs of fifteen antennas for three Raman bands at 1180 cm⁻¹, 1370 cm⁻¹ and 1618 cm⁻¹. Each three dots with the same color come from one same antenna. The dashed lines are the average EMEFs for each band.

According to Fig. 1e, the EMEFs of the three Raman bands at 1180, 1370 and 1618 cm$^{-1}$ are $10^{9.22\pm0.23}$, $10^{9.21\pm0.22}$ and $10^{9.22\pm0.24}$, respectively. The first number on the exponent is the average value of $Log_{10}$ EMEF, and the second number is the root-mean-square (RMS) error. These results show both ultrahigh enhancement and reasonable reproducibility. All of the twenty SERS spectra are shown in Supplementary Fig. S3. Further, by enclosing the antenna with 100 nm deep double rings to improve collecting the part of Raman scattering that couples to surface plasmon polaritons, higher EMEFs of the same three Raman bands have been measured for another fifteen antennas. The results are $10^{9.67\pm0.33}$, $10^{9.68\pm0.36}$ and $10^{9.67\pm0.36}$. The alignment error between the antennas and the double rings contribute to the EMEF errors. A scanning electron micrograph (SEM) of the antenna-with-double-ring device, a typical SERS spectrum, and the EMEF distribution are shown in Fig. 2. All of the fifteen SERS spectra are also shown in Supplementary Fig. S3. Excluding chemical enhancement, the SERS EFs in our experiments are comparable to the highest in previous reports on random aggregates[3,7,10,11]. In addition, thanks to the low laser power, the SERS signals were stable for more than five minutes without any obvious evidence of molecule degradation.

In the above, we have used MGITC as the probe molecule due to its large resonant Raman scattering cross section at 633 nm, so that the Raman signals from the bare gold plane can be measured to calculate EMEF. We also measured a monolayer of non-resonant small molecules, 4-nitrobenzenthiol (4NBT), from twenty antennas, as shown in Fig. 3. Not only can we observe clear SERS signals from the –NO$_2$ stretching mode at 1336 cm$^{-1}$ under 300 nW laser power and 4 s integration time, but a considerably better reproducibility than that of MGITC which is $10^{\pm 0.08}$. We suppose the higher reproducibility to be the consequence of the molecules' better chemical stability when they are non-resonant with the laser[25] and having a larger number of smaller molecules in each hotspot.

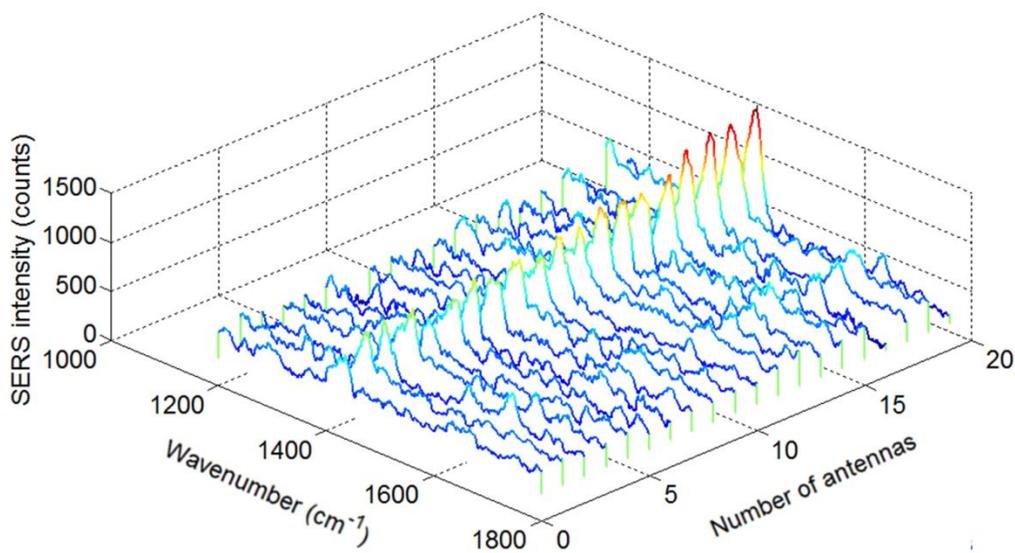

**Figure 3 | SERS of twenty antennas coated with a monolayer of 4NBT.** The laser power at sample is 300 nW. The integration time is 4 s. The background fluorescence spectra have been subtracted.

The value of $A_{\text{hotspot}}$ needs further discussion. Values from less than 1 nm$^2$ to several tens of nm$^2$ have been used in the literature. Ultra-small hotspots seem to be evidenced by ultrahigh resolution TERS mapping experiments, both under ultrahigh vacuum and low temperature and in ambient conditions[6,22]. The sub-nm TERS hotspots were related to the nonlinear dependence of TERS intensity on laser power, which was suggested to result from stimulated Raman scattering (SRS)[6]. In our experiment, a strong nonlinear dependence of the SERS intensity on the laser power has also been observed, as shown in Fig. 4. However, we can exclude the possibility of SRS effect by working with a low laser power, as explained in the following. 300 nW at 633 nm corresponds to $9.5 \times 10^{11}$ photons per second, and the plasmon life time in the antenna is 5.3 fs according to the LSPR bandwidth in Fig. 1b, so that no more than $5.0 \times 10^{-3}$ plasmons are simultaneously confined in the antenna under 300 nW laser power. Therefore SRS by the plasmons is much weaker than the spontaneous Raman scattering[26]. The origin of nonlinearity is not clear to us at the moment, and we don't know whether it implicates smaller hotspots and higher EFs than we have estimated in this paper.

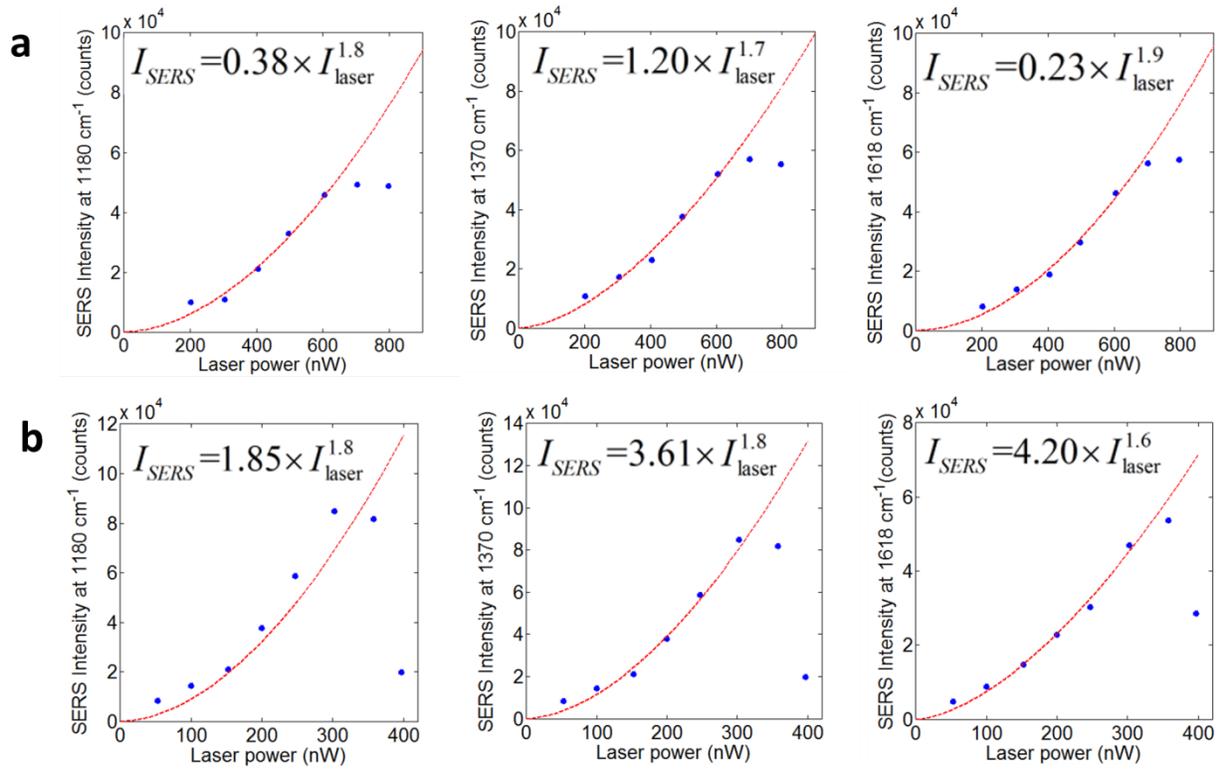

**Figure 3 | SERS versus laser power. a,** The SERS intensities of three Raman bands of MGITC versus laser power at sample, for an antenna coated with a monolayer of MGITC. **b,** The same as **a**, but for an antenna enclosed by double rings. The blue dots are measurement results. The red dashed curves and the equations in the figure are exponential fitting results. The integration time is 4 s.

Compared with previous work on nanoparticle-plane junction SERS, the enormously improved EFs and reproducibility in our experiment come from three factors. First, the RP laser beam is critical for obtaining maximum $|E_z|$ in the laser focal spot and consequently efficient excitation of the vertical antenna. In general, the laser beam and the focusing element should have the same central symmetry, while the other methods such as inclined illumination has been proved inefficient by us[27]. The vectorial profiles of both LP and RP focal spots are compared in Supplementary Fig. S4 to further illustrate this point. Second, the atomically flat gold plane is critical for reproducibility, while most previous work used evaporated or sputtered metal films which had nanometer scale surface roughness. The variance of EF in our experiment is largely because the nanospheres are actually polyhedrons with crystal plane facets, so that different nanospheres have different interfaces with the hotspots. A transmission electron micrograph (TEM) and LSPR spectra of twenty antennas are shown in Supplementary Fig. S5 for further discussion. Third, the EFs roll over at sub-µW laser powers, as shown in Fig. 4. The reproducibility significantly worsens in the roll-over regime, therefore low laser power operation is critical, which in turn requires high sensitivity.

In conclusion, by focusing an RP laser beam onto the gold nanosphere - atomically flat gold plane junctions, we have obtained ultrahigh SERS EFs that are quite uniform between different hotspots. Together with the benefits of low power operation, this method should facilitate systematic study of nanoscale molecular behavior by Raman spectroscopy[28]. It also provides a sensitive and reproducible probe for exploring the physics of nanoscale hotspots, e.g. nonlinearity[6], nonlocality and quantum tunneling[16,29,30]. In the future, we will integrate the nanosphere with the tip of a scanning force microscope for imaging and precise gap size control.

## Methods

**Sample preparation:**

The antennas coated with a monolayer of MGITC were prepared as follows[17]. First, 13.5 μL of 45 μM MGITC (Invitrogen M689) ethanol solution and 1 mL of $5.2 \times 10^9$/mL gold nanosphere ultra-purified water solution (BBI Solutions, 60 nm mean diameter, ±8% variation) were incubated together for 2 hours at room temperature. Then the functionalized gold nanosphere solution was 1:1 diluted with ultra-purified water, and drop-casted onto the gold planes. The gold planes are 200 nm thick Au (111) films on mica substrates (PHASIS), which have been deposited by magnetron sputtering and then hydrogen flame annealed to obtain atomically flat surfaces. Next the samples were rinsed with ultra-purified water and dried under a stream of nitrogen.

The bare gold planes coated with a monolayer of MGITC were prepared as follows. The gold plane samples were first immersed in 1 μM MGITC ethanol solutions for 10 minutes, then rinsed with ethanol and dried under a stream of nitrogen[24].

The antennas coated with a monolayer of 4NBT were prepared as follows. First, 0.5 mL of $6.5 \times 10^9$ /mL gold nanosphere ultra-purifed water solution was added to 0.5 mL of 4 μM 4NBT (Sigma-Aldrich) water solution and mixed for 2 hours at room temperature. Then the functionalized gold nanosphere solution was drop-casted onto the gold planes. Next the samples were dried under a stream of nitrogen.

Dual-beam focused ion beam (FIB) milling was used to fabricate the double rings and align the rings' centers to the nanospheres.

**Optical measurement**:

Raman scattering was measured as follows. A He-Ne laser working at 632.8 nm and $TEM_{00}$ mode was used to excite the molecules. The laser beam passed through a liquid crystal polarization converter (ARCoptix) and was converted to an RP state of polarization. The RP laser beam was focused onto the samples through a long working distance $100 \times$ Plan Apo objective, whose NA is 0.9. Reflection from the sample, including Raman scattering, was collected by the same objective, passed through a long-pass filter, and detected by a monochromator installed with an electron multiplying CCD (EMCCD) detector.

LSPR was measured as follows. A super continuum source was focused onto the samples through the same $100 \times$ objective. The scattered light was collected outside the NA of the objective with a lens whose NA is 0.15. The collecting lens focused the scattered light into a fiber-bundle, which was directed to the monochromator and EMCCD detector. The power of the super continuum source was carefully decreased by neutral density filters in order not to damage the samples.

The small scattering cross section of the antennas and the large reflection off the gold plane render it extremely difficult to find the nanospheres under optical microscopes without special methods. The same $100\times$ objective was used as part of a home-built microscope to observe the nanospheres. A spatial filter blocked the central part of the objective's entrance pupil so that the nanospheres were illuminated at an inclined angle. The nanospheres appear as dark spots on a bright background, due to the antennas' absorption and scattering of the inclined illumination. In addition, FIB milled position markers were made on the gold planes, and SEM images were taken to compare with the optical microscopy images, so that the nanospheres can be identified repeatedly.

We have selected those gold nanoparticles with spherical shapes under SEM for optical experiments. Around 10% of the gold nanoparticles have irregular non-spherical shapes. Otherwise, we have not intentionally excluded any nanospheres for SERS EF reproducibility characterization.

**FDTD simulation**

The FDTD simulations were done with Lumerical FDTD Solutions. The nanosphere-plane junction structure is excited by a broadband total-field scattered-field source, which is a *p*-polarized planewave at 30°-to-normal incidence. The boundary conditions are perfectly matched layers except for one mirror symmetry plane across the center of sphere. The finest grid size of the mesh is 0.05 nm in and near the junction gap, and increases to 4 nm at away from the junction gap.

## Acknowledgements

The authors thank B. Ren and X. Wang from Xiamen University, Q. Zhan from University of Dayton, and S. Guo and W. Shen from Shanghai Jiao Tong University for useful advices. This work is supported by the National Science Foundation of China under grant # 11204177 and # 11574207, the Fundamental Research Program of Science and Technology Commission of Shanghai Municipality under grant # 14JC1491700, the Research Fund for the Doctoral Program of Higher Education of China under grant # 20120073110050, and the Center for Advanced Electronic Materials and Devices of Shanghai Jiao Tong University.

## Author contributions

J.L. and T.Y. designed the experiments and analyzed the data. J.L. performed all the experiments. H.Y. conducted all the numerical simulations and the focal spot profile calculation. All authors contributed to building the optical measurement system. T.Y. and J.L. wrote the paper.

# Reproducible Ultrahigh Electromagnetic SERS Enhancement in Nanosphere-Plane Junctions

Jing Long, Hui Yi, Hongquan Li and Tian Yang

## 1. Hotspot intensity spectrum by FDTD simulation

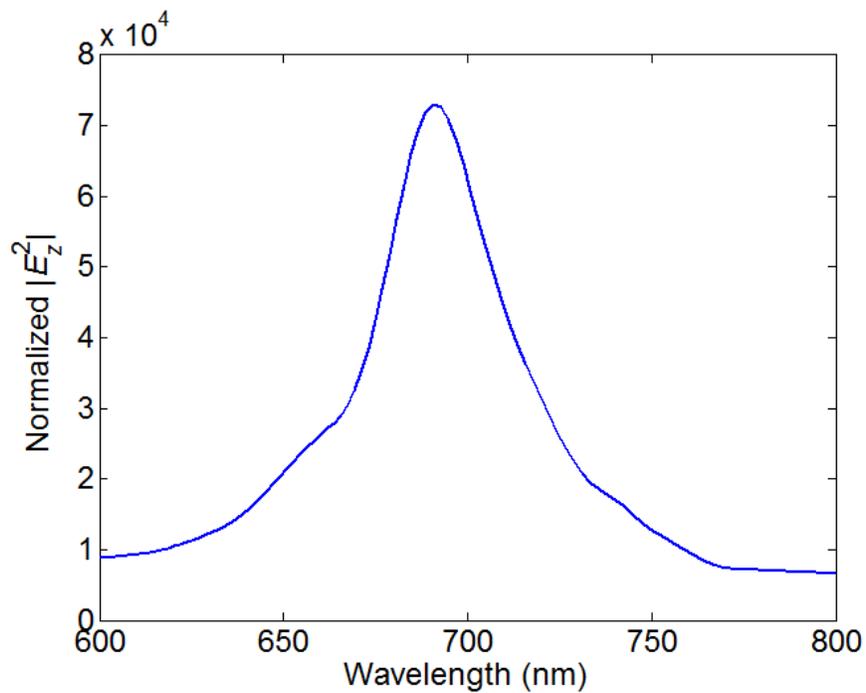

**Figure S1 FDTD simulation of $|E_z^2|$ at the center of the antenna's junction gap.** The antenna in Fig. 1. is used. The illumination is an around 30°-to-normal $p$-polarized broadband planewave. The intensity is normalized by the $|E_z^2|$ of illumination.

## 2. Supplementary Methods: EMEF calculation

The EMEFs of our experiments are calculated by equation (1).

$$EMEF = \frac{I_{\text{SM-hotspot}}}{I_{\text{SM-air}}}$$

$$= \frac{P_{\text{laser2}}}{P_{\text{laser1}}} \cdot \frac{I_{\text{hotspot}} / A_{\text{hotspot}}}{I_{\text{plane}} / A_{\text{RP}}} \cdot 2^4 \cdot \frac{\left| E_{\text{RP}}{}^2 \right|}{\left| E_{\text{LP}}{}^2 \right|} \tag{1}$$

The first line of equation (1) is the definition of SERS EF, or the definition of EMEF when the chemical contribution to EF is excluded as in this paper. $I_{\text{SM-hotspot}}$ is the collected Raman scattering from a single molecule in the junction gap hotspot. $I_{\text{SM-air}}$ is the collected Raman scattering in an imaginary experiment where an LP laser beam with the same power is focused onto a single molecule in air, using the same focusing and collecting objective as in our experiment. The second line is how we calculate the EMEF. $P_{\text{laser1}}$ is the laser power for exciting the molecules in the hotspot, which is 300 nW. $P_{\text{laser2}}$ is the laser power for exciting the molecules on the bare gold plane, which is 1.5 mW. $I_{\text{hotspot}}$ and $I_{\text{plane}}$ are the collected Raman scattering from a monolayer of molecules in the hotspot and on the bare gold plane divided by their respective integration time. $A_{\text{hotspot}}$ and $A_{\text{RP}}$ are the full-width-half-maximum (FWHM) area of the hotspot and the laser focal spot, respectively, so that $I/A$ is proportional to Raman scattering power per molecule. Note that $A_{\text{hotspot}}$ is the area of $|E^4|$ since SERS EMEF in an LSPR hotspot is proportional to $|E^4|$, while $A_{\text{RP}} = 0.08\ \mu m^2$ is the area of $|E^2|$ since Raman is a linear process by itself [1,2]. Due to the mirror effect of the gold plane, the transverse $E$ is weak and only $E_z$ is considered in the estimation of $A$'s. The $2^4$ factor counts for the EM enhancement contributed by the bare gold plane compared to the imaginary in-air experiment, which includes the mirror effect which increases the excitation $|E_z{}^2|$ by a factor of around $2^2$, and the Purcell effect which increases the Raman emission rate by another factor of around $2^2$. $|E_{\text{RP}}{}^2|/|E_{\text{LP}}{}^2|$ equals the ratio between the $|E^2|$ values at the center points of the RP and LP focal spots, which is theoretically calculated to be 0.89. A line-scanning of the FDTD simulation result of the hotspot (Fig. 1c) is shown in Fig. S2, which gives a FWHM hotspot diameter of 3.45 nm and $A_{hotspot}$ of 9.3 nm$^2$.

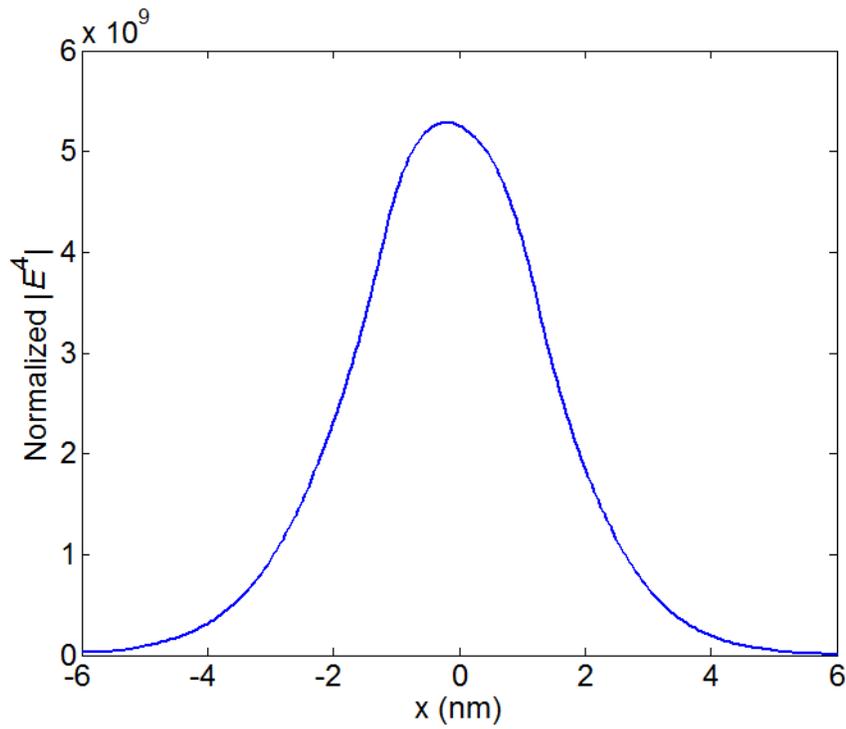

**Figure S2 Line scanning of $|E^4|$ across the center of the hotspot.** FDTD simulation result is plotted. The $|E^4|$ value has been normalized to the incident field.

## 3. SERS of multiple antennas coated with monolayer MGITC

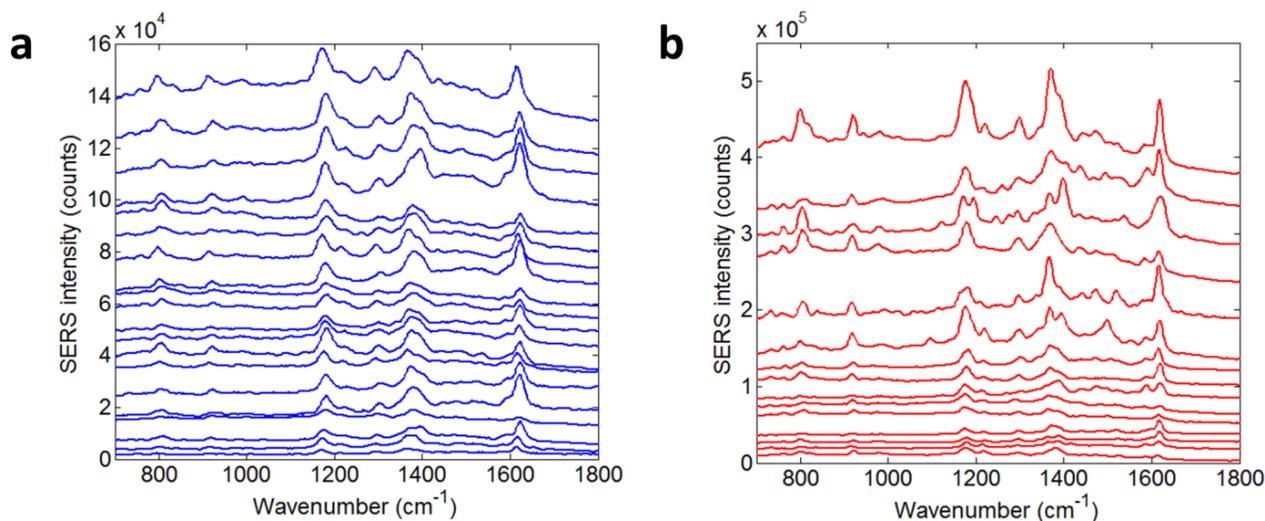

**Figure S3**. **SERS spectra of multiple antennas coated with a monolayer of MGITC**. The laser power at sample is 300 nW. The integration time is 4 s. Each spectrum has been lifted to a different height in the figures. **a,** Twenty antennas. **b,** Fifteen antennas with double rings

## 4. Comparison between RP and LP focal spots

RP and LP laser beams pointing in the $z$ direction are focused through an objective with NA=0.9. The vectorial profiles of the focal spots are experimentally characterized by raster scanning a gold nanosphere on a silica aerogel substrate and measuring the scattered far field, following Ref [3]. The theoretical calculation results are also presented, following Ref [4]. The vectorial profiles shown in Fig. S4 indicate much larger $|E_z^2|$ in the RP focal spot than in the LP focal spot.

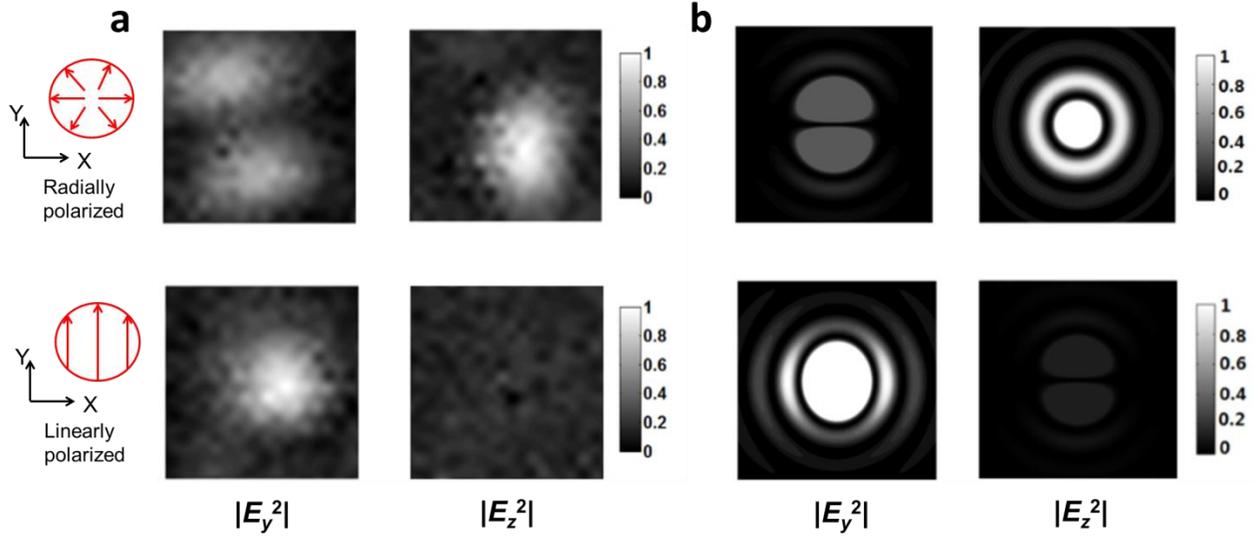

**Figure S4 Vectorial profiles of RP and LP focal spots.** The first row is RP focal spot profiles. The second row is LP focal spot profiles. Each column is the profile of either $|E_y{}^2|$ or $|E_z{}^2|$, as labeled. Laser wavelength is 632.8 nm. Focusing objective NA=0.9. All images are 1.2×1.2 μm². **a**, Experimental results. **b**, Theoretical results. The intensities are normalized to the maximum $|E^2|$ of the respective focal spots.

## 5. Effects of nanosphere-plane interfaces

The TEM image of a 60 nm gold nanosphere coated with a monolayer of MGITC is shown in Fig. S5a. It is a polyhedron. The contact between the polyhedron and the plane may be in the form of facet, edge or apex. We have observed significant deviation between the LSPR spectra of different antennas, as shown in Fig. S5b and c. Three out of twenty antennas show double peaks in their LSPR spectra, as the orange curve in Fig. S5c. This is reported to result from strong charge concentration and plasmon coupling at the junction, which is sensitive to the morphology of the interface [5,6]. On the other hand, the high reproducibility of SERS EFs implicates that the values of $A_{\text{hotspot}}$ may not vary a lot between different antennas.

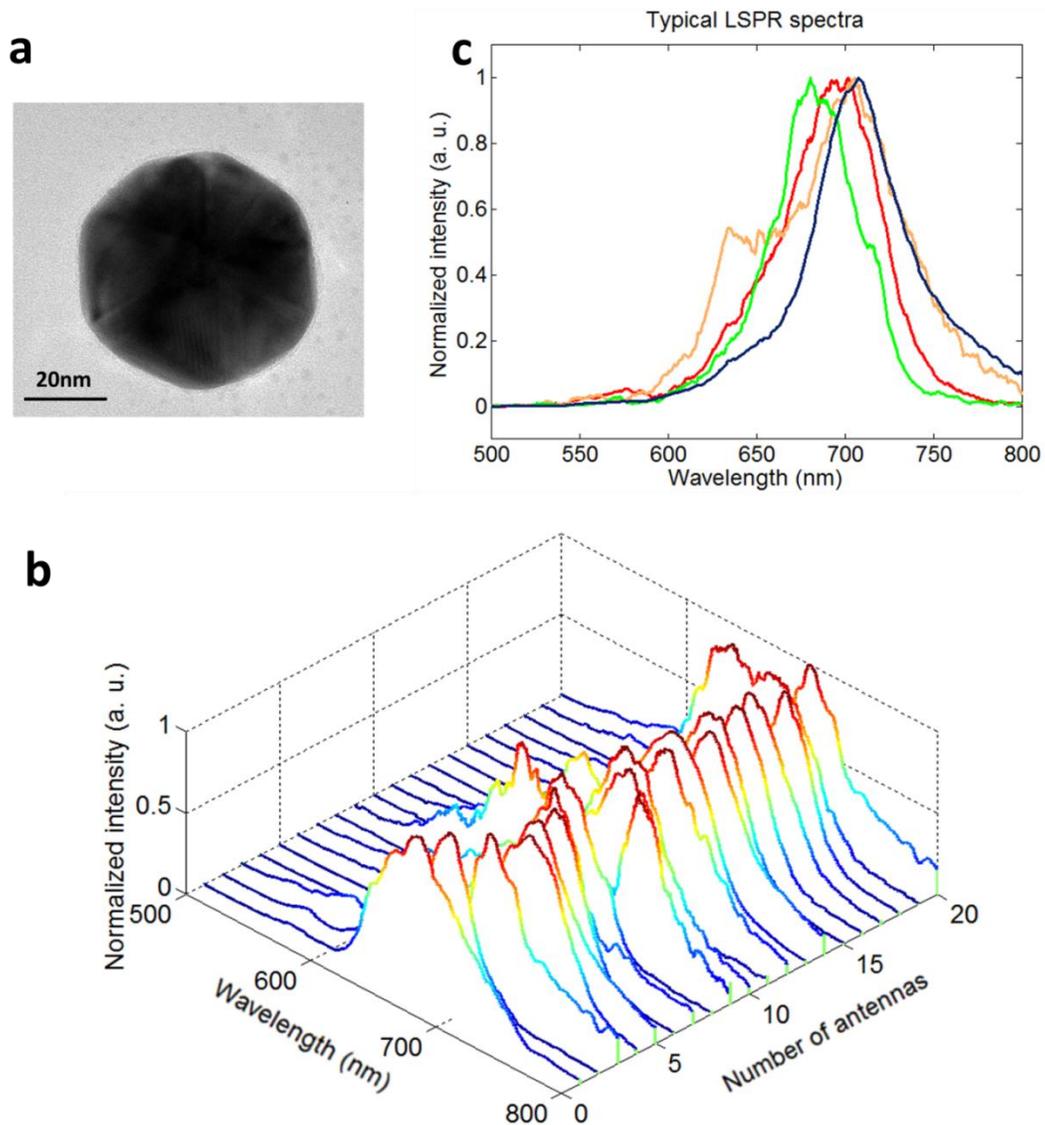

**Figure S5 a**, TEM image of a 60 nm gold nanosphere coated with a monolayer of MGITC. **b**, LSPR spectra of twenty antennas, normalized to the same height. These antennas are the same as in Fig. S3a. **c**, Some representative LSPR spectra from **b**.